\documentclass[journal=ancac3,manuscript=article,layout=twocolumn]{achemso}

\usepackage{chemformula} 
\usepackage[T1]{fontenc} 
\usepackage{hyperref}    

\author{Liam Watson}
\affiliation[Monash University]{School of Physics and Astronomy, Monash University, Clayton, 3800, VIC, Australia}
\alsoaffiliation{Australian Research Council Centre of Excellence in Future Low-Energy Electronics Technologies, Monash University, Clayton, 3800, VIC, Australia}

\author{Joan Ripoll}
\affiliation[IMDEA]{Instituto Madrile\~{n}o de Estudios Avanzados en Nanociencia (IMDEA-Nanociencia), Madrid, 28049 Spain}
\alsoaffiliation{Departamento de F\'{i}sica de la Materia Condensada, Universidad Aut\'{o}noma de Madrid, Madrid, 28049 Spain}

\author{Zhengjue Tong}

\author{Amit Kumar}

\author{Yande Que}
\affiliation[NTU]{School of Physical and Mathematical Sciences, Nanyang Technological University, Singapore, 637371 Singapore}

\author{Yang-Hao Chan}
\affiliation[Academia Sinica Sciences]{Institute of Atomic and Molecular Sciences, Academia Sinica, Taipei 106319, Taiwan}

\author{Hsin Lin}
\affiliation[Academia Sinica Physics]{Institute of Physics, Academia Sinica, Taipei 155201, Taiwan}

\author{Shantanu Mukherjee}
\affiliation[Madras]{Quantum Centre for Diamond and Emergent Materials, Indian Institute of Technology Madras, Chennai, Tamil Nadu 600036, India}

\author{Manuela Garnica}
\affiliation[IMDEA]{Instituto Madrile\~{n}o de Estudios Avanzados en Nanociencia (IMDEA-Nanociencia), Madrid, 28049 Spain}
\alsoaffiliation{Instituto Nicolás Cabrera, Universidad Autónoma de Madrid,
28049 Madrid, Spain}

\author{Mark T Edmonds}
\affiliation[Monash University]{School of Physics and Astronomy, Monash University, Clayton, 3800, VIC, Australia}
\alsoaffiliation{Australian Research Council Centre of Excellence in Future Low-Energy Electronics Technologies, Monash University, Clayton, 3800, VIC, Australia}

\author{Micha\l\ Papaj}
\affiliation[University of Houston]{Department of Physics, University of Houston, Houston, Texas 77204, USA}

\author{Amadeo L Vazquez de Parga}
\affiliation[IMDEA]{Instituto Madrile\~{n}o de Estudios Avanzados en Nanociencia (IMDEA-Nanociencia), Madrid, 28049 Spain}
\alsoaffiliation{Departamento de F\'{i}sica de la Materia Condensada, Universidad Aut\'{o}noma de Madrid, Madrid, 28049 Spain}
\alsoaffiliation{IFIMAC Condensed Matter Physics Center, Madrid, 28049, Spain}

\author{Bent Weber}
\email{b.weber@ntu.edu.sg}
\affiliation[NTU]{School of Physical and Mathematical Sciences, Nanyang Technological University, Singapore, 637371 Singapore}
\alsoaffiliation{Australian Research Council Centre of Excellence in Future Low-Energy Electronics Technologies, Monash University, Clayton, 3800, VIC, Australia}

\author{Iolanda Di Bernardo}
\email{iolanda.dibernardo@uam.es}
\affiliation[Monash University]{School of Physics and Astronomy, Monash University, Clayton, 3800, VIC, Australia}
\alsoaffiliation{Australian Research Council Centre of Excellence in Future Low-Energy Electronics Technologies, Monash University, Clayton, 3800, VIC, Australia}
\alsoaffiliation{Departamento de F\'{i}sica de la Materia Condensada, Universidad Aut\'{o}noma de Madrid, Madrid, 28049 Spain}
\alsoaffiliation{IFIMAC Condensed Matter Physics Center, Madrid, 28049, Spain}

\author{Michael S Fuhrer}
\email{michael.fuhrer@monash.edu}
\affiliation[Monash University]{School of Physics and Astronomy, Monash University, Clayton, 3800, VIC, Australia}
\alsoaffiliation{Australian Research Council Centre of Excellence in Future Low-Energy Electronics Technologies, Monash University, Clayton, 3800, VIC, Australia}

\title[Excitonic order parameter in WTe$_2$]{Observation of charge density wave excitonic order parameter in topological insulator monolayer WTe$_2$}

\abbreviations{FT-STS, QPI, CDW, SDW, LDOS, HOPG, STS, CNP, CG, STM, MBE, ARPES}

\begin{document}

\begin{tocentry}

\includegraphics{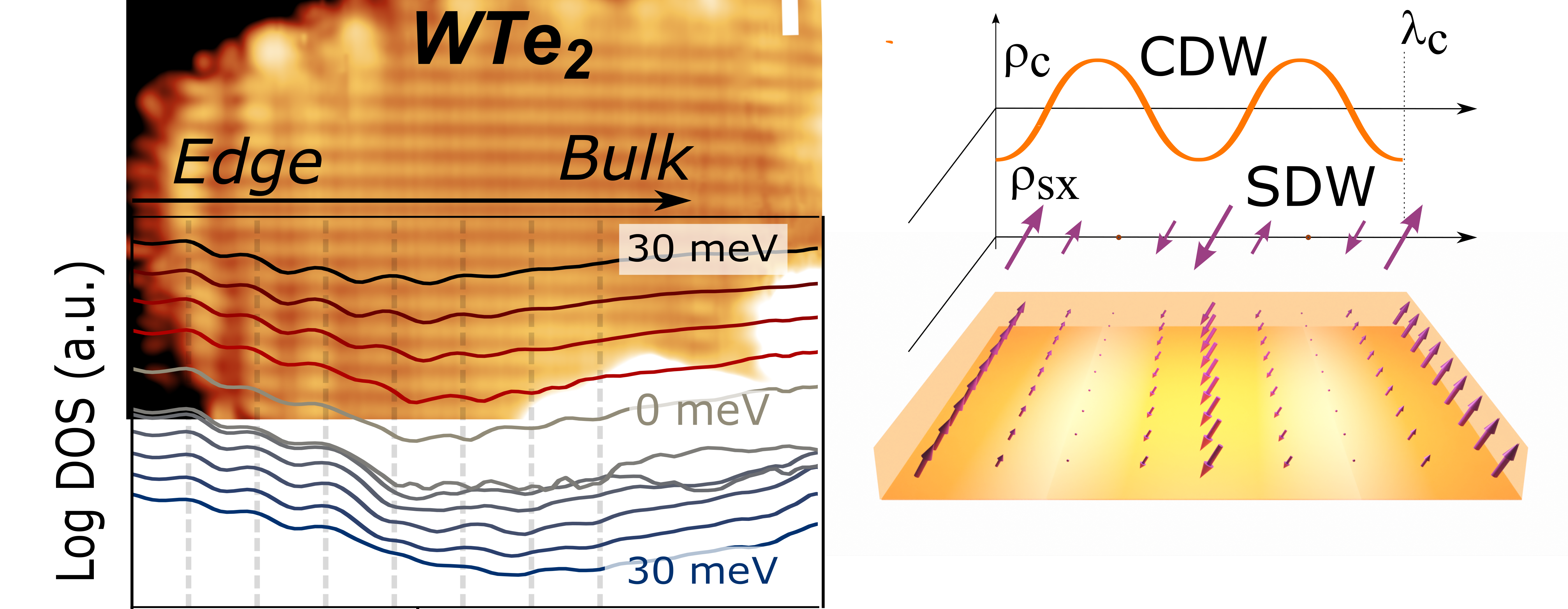}





\end{tocentry}


\begin{figure*}[t!]
  \centering
  \includegraphics[width = \textwidth]{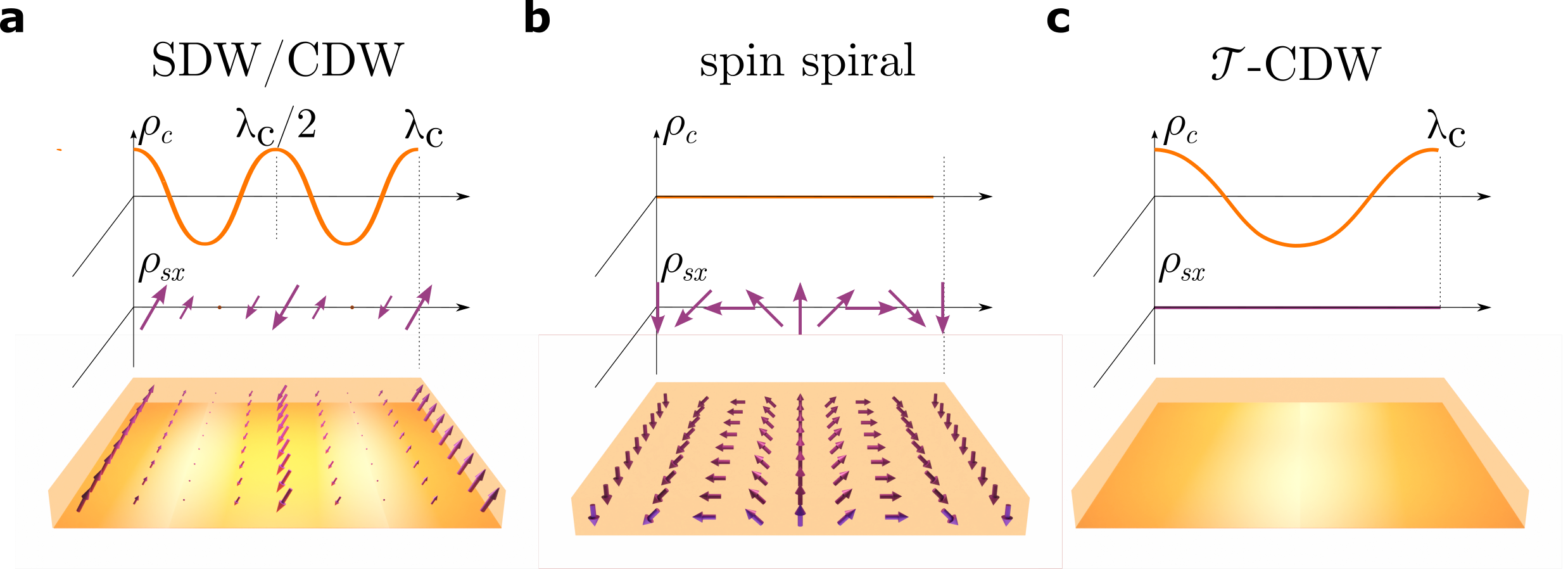}
  \caption{Schematic of the excitonic phases in WTe$_2$. The real space orders are defined in terms of the wavelength $\lambda_c = 1/q_c$.~\textbf{a.} SDW/CDW phase, exhibiting a charge density ($\rho_c$) with period $\lambda_c/2$ and spin density ($\rho_{sx}$) with period $\lambda_c$.~\textbf{b.} Spin spiral phase \textendash\ lacking both charge and spin density \textendash\ exhibits a spin rotation about the $y$-axis with period $\lambda_c$.~\textbf{c.} $\mathcal{T}$-CDW phase, exhibiting charge order with period $\lambda_c$ and lacks spin order.}
  \label{fig1}
\end{figure*}

\begin{abstract}
  Strong electron-hole interactions in a semimetal or narrow-gap semiconductor may drive a ground state of condensed excitons. Monolayer WTe$_2$ has been proposed as a host material for such an exciton condensate, but the order parameter \textendash\ the key signature of a macroscopic quantum-coherent condensate \textendash\ has not been observed. Here we use Fourier-transform scanning tunnelling spectroscopy (FT-STS) to study quasi-particle interference (QPI) and periodic modulations of the local density of states (LDOS) in monolayer WTe$_2$. In WTe$_2$ on graphene, in which the carrier density can be varied \textit{via} back-gating, FT-STS shows QPI features in the 2D bulk bands, confirming the interacting nature of the bandgap in neutral WTe$_2$ and the semi-metallic nature of highly \textit{n}- and \textit{p}-doped WTe$_2$. We observe additional non-dispersive spatial modulations in the LDOS imprinted on the topological edge mode of neutral WTe$_2$ on metallic substrates (graphene and graphite), which we interpret as the interaction of the topological edge mode with the expected charge density wave order parameter of the excitonic condensate in WTe$_2$ at low interaction strength due to screening by the metallic substrates. 
\end{abstract}

\section{Introduction}

\paragraph{}
In a semimetal or narrow-gap semiconductor the Coulomb interaction can drive a unique many-body ground state in which excitons \textendash~composite bosons of strongly bound electrons and holes \textendash~condense. The exciton condensate is a macroscopic quantum-coherent state, analogous to a superconductor, with a BCS-like order parameter~\cite{cloizeaux_EITheory_1965,kozlov_EITheory_1966,Jerome1967_Excitonic_Insulator,halperin_EITheory_1968}. However because excitons are charge-neutral, the condensate allows no supercurrent and is an example of an excitonic insulator. The existence and behaviour of the exciton condensate is highly conditional on the electron and hole densities and the strength of the Coulomb interaction ($U$), and so is expected to depend strongly on doping, electric fields, and screening from the dielectric environment (where $U \propto \varepsilon^{-1}$; $\varepsilon$ is the dielectric constant). Monolayers of WTe$_2$ are theoretically predicted to host an exciton condensate with at least three possible phases which will be referred to as SDW/CDW, spin spiral, and $\mathcal{T}$-CDW respectively. In the time-reversal symmetry breaking SDW/CDW phase at low $U$, the order parameter exhibits a spin density wave (SDW) with wavevector $q_c$ and a charge density wave (CDW) with wavevector $2q_c$~\cite{Papaj_CDW_QPI_2024,Kwan_SSvsSDW_2020}, where ${q_c}$ is expected to be roughly  equal to the separation of the electron and hole pockets in momentum space. At high values of $U$, however, the condensate is expected to exhibit a time-reversal breaking spin spiral phase, which is absent of both charge and spin density modulations, but these can be weakly revealed upon application of a magnetic field~\cite{Kwan_SSvsSDW_2020,Cobden2021_WTe2_EI}. A time-reversal symmetry preserving phase $\mathcal{T}$-CDW  has been described, which is nearly degenerate in energy to the spin spiral phase, and exhibits a CDW order parameter, but with half the wavevector ($q_c$) and is absent of any spin order~\cite{Papaj_CDW_QPI_2024,Wang_breakdown_2023}. 

\paragraph{}
Recent experimental evidence~\cite{Bent2024_WTe2_EI_Gating,Jia2021_WTe2_EI_Transport,Cobden2021_WTe2_EI} suggests monolayer WTe$_2$ to be a topological excitonic insulator, with conducting topological edge modes and an insulating bulk which is insensitive to charge addition~\cite{Jia2021_WTe2_EI_Transport,Cobden2021_WTe2_EI} up to a critical threshold at which the gap collapses~\cite{Bent2024_WTe2_EI_Gating}. However, direct evidence for an exciton \textit{condensate} in WTe$_2$, \textit{e.g.}~the existence of a symmetry-breaking order parameter indicating spontaneous macroscopic coherence, is lacking. The failure to observe the order parameter in the WTe$_2$ ground state is not understood, but it has been speculated to be due to the difficulty of detecting the spin spiral phase with no charge ordering, which is expected in some experimental circumstances~\cite{Kwan_SSvsSDW_2020,Jia2021_WTe2_EI_Transport}. CDW-like oscillations have been observed in early reports of monolayer WTe$_2$~\cite{Jia2017_WTe2_QPI_STM, Song2018_WTe2_QPI_STM}, but those samples displayed a semimetal-like local density of states (LDOS) with a Coulomb gap centred at the Fermi energy. Fourier transform scanning tunnelling spectroscopy (FT-STS) spectra on those samples revealed quasiparticle interference (QPI) features from overlapping electron and hole pockets, and so the possibility of charge order was rejected in favour of interpocket scattering around the Fermi level~\cite{Song2018_WTe2_QPI_STM}. Recently, evidence of edge LDOS modulations of order $q_c$ were observed in samples of monolayer WTe$_2$ on graphene~\cite{Bent2024_WTe2_EI_Gating,Jia_WTe2_STM_Luttinger2022}, although a systematic analysis of their dispersion was not undertaken. 


\begin{figure*}[t!]
  \centering
  \includegraphics[width = \textwidth]{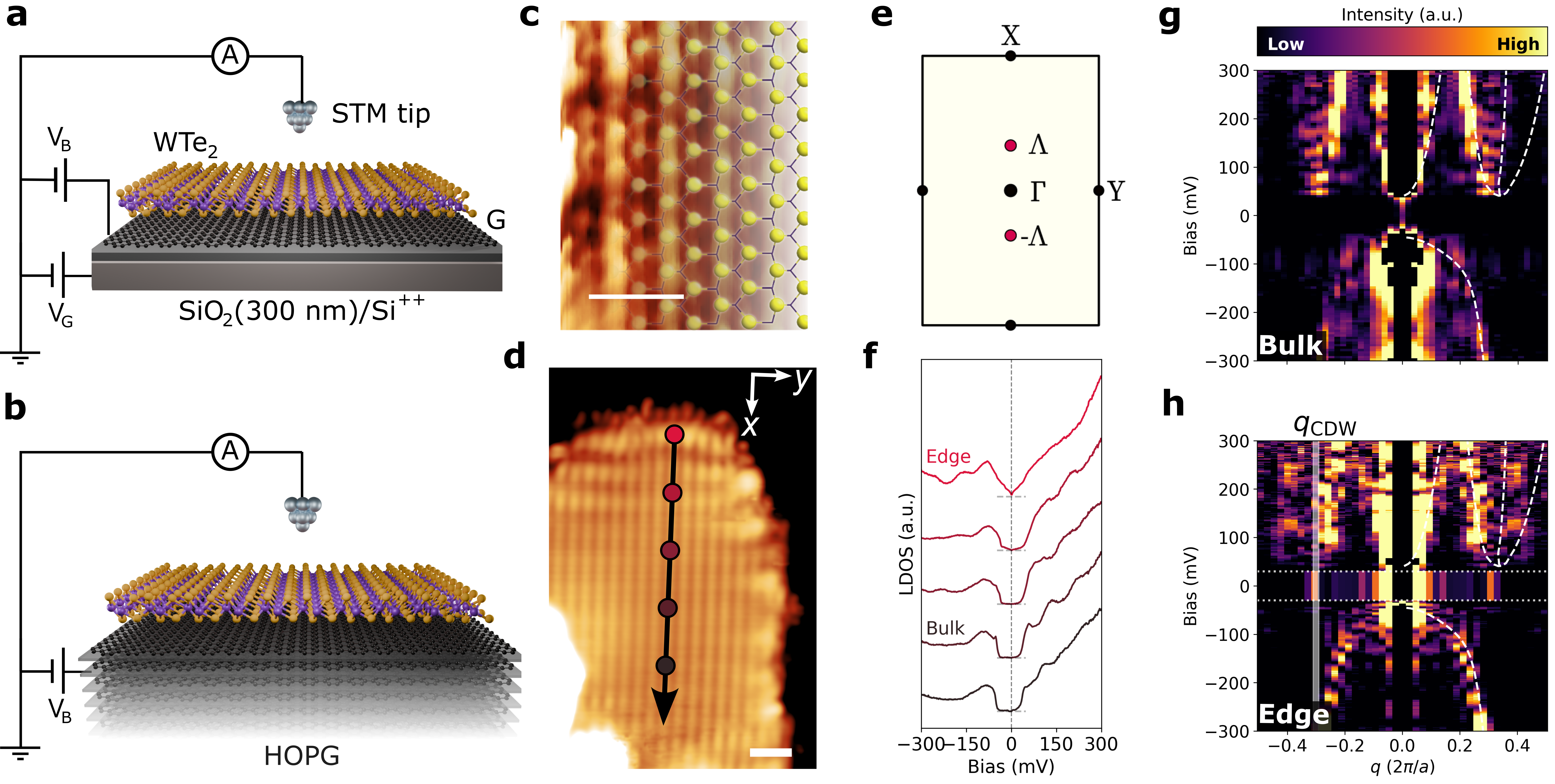}
  \caption{Experimental overview.~\textbf{a,b.} Schema of STM experimental setups. Epitaxial sub-monolayer WTe$_2$ grown on \textbf{a.} graphene (G) over SiO$_2$/Si, and \textbf{b.} HOPG\@. Sample ($V_B$) and back-gate ($V_G$) electrode biases are indicated.~\textbf{c.} Topography of the WTe$_2$ surface, overlaid with the atomic lattice. Scale bar = 1 nm.~\textbf{d.} Topography of WTe$_2$ on HOPG\@. Lattice directions are indicated. Scale bar = 2 nm.~\textbf{e.} First Brillouin zone of WTe$_2$. The high symmetry points ($\Gamma$, X, Y) and electron hole pockets $\pm \Lambda$ are indicated.~\textbf{f.} LDOS measured along the black arrow in d at the points indicated.~\textbf{g,h.} FT-STS spectra for \textbf{g.} bulk, and \textbf{h.} edge regions of WTe$_2$. The gap energies (between $-30$ and $+30$ meV) in h are integrated for clarity. The calculated scattering vectors from the interacting $k \cdot p$ band structure and the measured CDW wavevector are indicated with white dashed lines and the white shaded region respectively. STM/STS parameters: \textbf{c,d.} $V = 50$ mV, $I = 100$ pA, \textbf{g.} $V_\text{set} = 500$ mV, $I_\text{set} = 2$ nA, $V_\text{mod} = 5$ mV, $f_\text{mod} = 726$ Hz.~\textbf{f,h.} $V_\text{set} = 300$ mV, $I_\text{set} = 1$ nA, $V_\text{mod} = 2.3$ mV, $f_\text{mod} = 736$ Hz.}
  \label{fig2}
\end{figure*}

\paragraph{}
Here we investigate the electronic structure of monolayer WTe$_2$ using FT-STS\@. The topological nature of WTe$_2$ is confirmed \textit{via} the observed gapless edge mode. We resolve QPI features characteristic of the insulating bulk band structure, with a single valence band maximum at zero wavevector $\Gamma$ and two conduction band minima at wavevector $\Lambda \approx q_c$, separated by an energy gap. Charge injection \textit{via} gating of the WTe$_2$ causes the energy gap to close at critical electron and hole densities, whereupon QPI directly confirms that the collapsed-gap states are semimetals. The QPI features are in good agreement with calculated interacting $k \cdot p$ model band structure. Near the sample edges we observe similar QPI features but also additional modulations of the LDOS which decay from the edge into the bulk, that are non-dispersive, persist through the bandgap, and do not originate from QPI\@. We interpret these LDOS modulations as the signature of the charge density wave exciton condensate order parameter at $2q_c$, imprinted on the topological edge state, confirming the quantum coherent nature of the insulating state.

\section{Results}

\paragraph{}
\autoref{fig2} shows a summary of our experimental evidence for an exciton condensate order parameter in WTe$_2$. Monolayer WTe$_2$ was grown \textit{via} molecular beam epitaxy (MBE) on two different sample architectures shown in~\autoref{fig2}a,b: exfoliated graphene on 300 nm SiO$_2$ over \textit{p}-type silicon, and highly-oriented pyrolytic graphite (HOPG), respectively. Graphite and graphene were chosen to ensure negligible substrate coupling effects~\cite{tao_WTe2_graphiticSubstrateCoupling_2022} and high conductivity at ultra-low temperatures. The carrier concentration in the graphene was tuned by application of a voltage $V_G$ to the \textit{p}-type silicon back-gate. Charge transfer between the graphene and epitaxial WTe$_2$ is driven by the generated chemical potential offset, facilitating electrostatic doping~\cite{Bent2024_WTe2_EI_Gating}. The crystallisation in a rectangular 1T$^\prime$ lattice ($a = 3.57$ \r{A} and $b = 6.25$ \r{A}) was confirmed by low-energy electron diffraction (not shown) and atomically resolved STM images (\autoref{fig2}c).

\paragraph{}
Scanning tunnelling spectroscopy (STS), which measures the energy dependent LDOS, was performed at points along a line from the monolayer edge into the bulk along the \textit{x}-direction indicated in~\autoref{fig2}d, which corresponds to the cut along $\Gamma$--X in the Brillouin zone, intersecting both the hole pocket (centred at $\Gamma$) and electron pockets (centred at $\pm \Lambda$) (\autoref{fig2}e).~\autoref{fig2}f shows that the insulating interior is confirmed in STS \textit{via} the observation of a hard gap of width $\sim 60$ meV roughly centred around the Fermi energy, while the enhanced and gapless LDOS at the edge confirm the topological properties of the monolayer \textit{via} the bulk-boundary correspondence.

\begin{figure}[p]
  \centering
  \includegraphics[width=\linewidth]{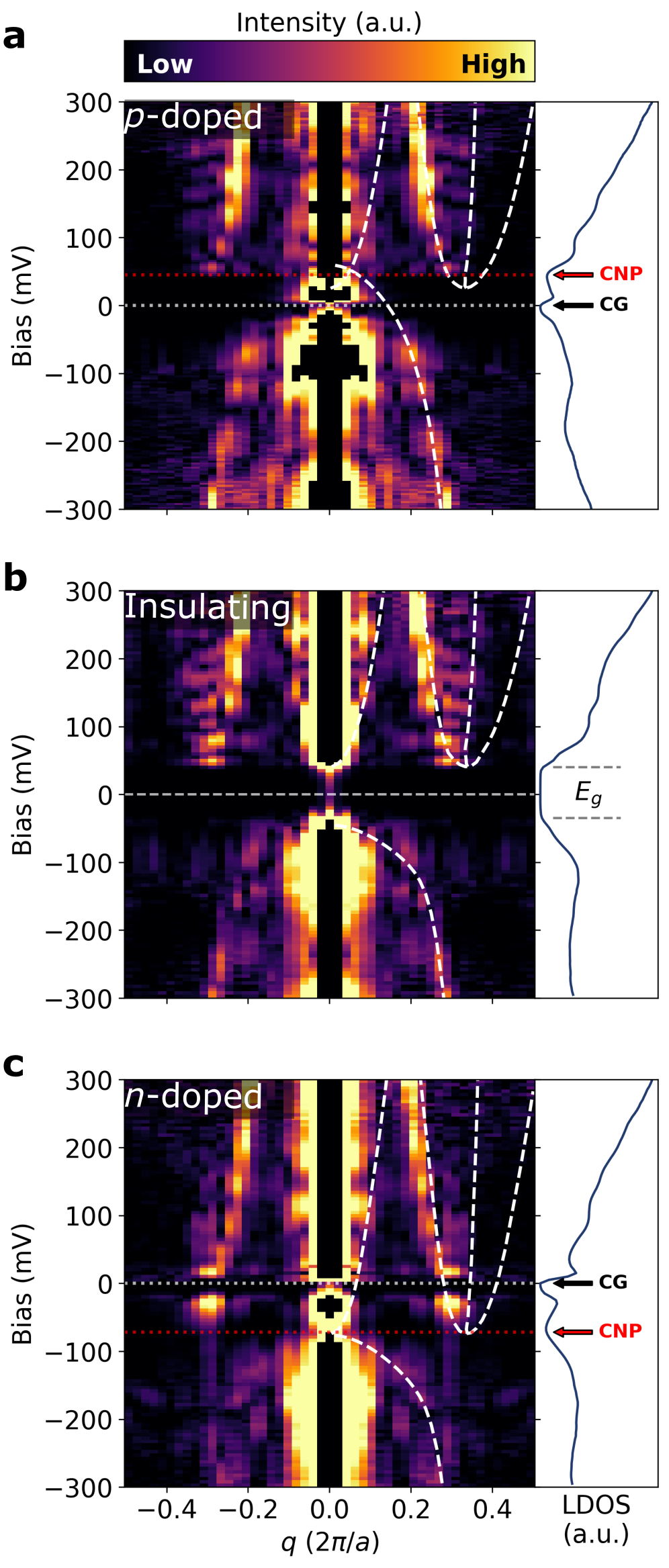}
  \caption{FT-STS spectra from the monolayer WTe$_2$ bulk for various back-gate potentials: \textbf{a.} \textit{p}-doped ($V_G = -40$ V), \textbf{b.} insulating ($V_G = 10$ V), and \textbf{c.} \textit{n}-doped ($V_G = +70$ V). The calculated scattering vectors from the interacting $k \cdot p$ band structure are indicated with white dashed lines. Each FT-STS spectrum is accompanied by a corresponding integrated LDOS spectrum. The bandgap ($E_g$), charge neutrality point (CNP) and Coulomb gap (CG) are indicated. STS parameters: $V_\text{set} = 500$ mV, $I_\text{set} = 2$ nA, $V_\text{mod} = 5$ mV, $f_\text{mod} = 726$ Hz.}
  \label{fig3}
\end{figure}

\begin{figure*}[t!]
  \centering
  \includegraphics[width = \textwidth]{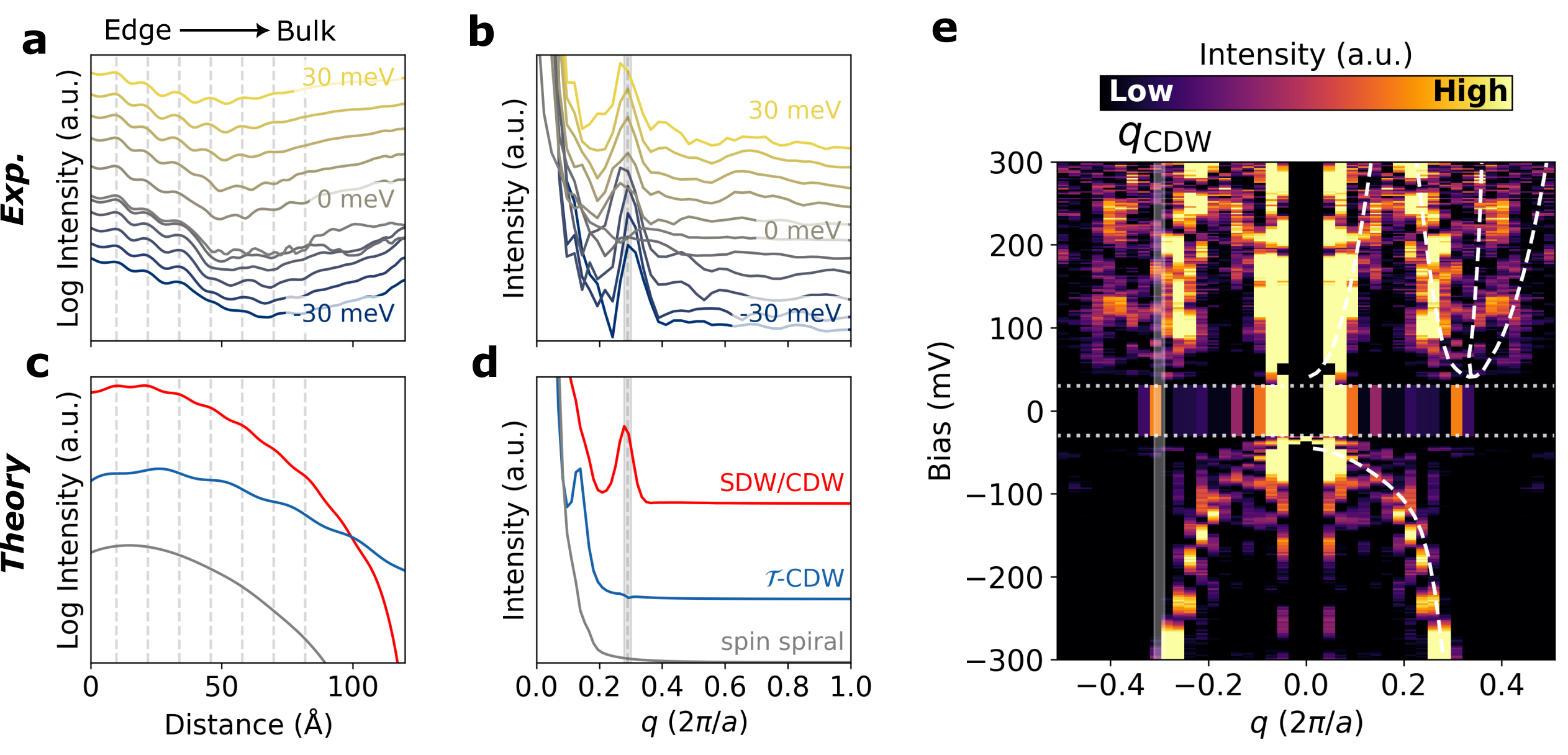}
  \caption{LDOS oscillations coupled to the topological edge state.~\textbf{a.} Constant energy cuts of the LDOS inside the insulating gap as a function of distance away from the edge, aquired at $B_\perp =$ 3 T.~\textbf{b.} Fourier transform of a.~\textbf{c.} Calculated topological edge LDOS and \textbf{d.} Fourier transform for the three different excitonic phases: SDW/CDW (red), $\mathcal{T}$-CDW (blue), and spin spiral (grey), offset for clarity. The approximate oscillation period in real-space is indicated with vertical dashed lines in a and c. The mean peak frequency and resolution in Fourier space are indicated by the grey dashed line and shaded region in b and d.~\textbf{e.} FT-STS spectrum of edge region of monolayer WTe$_2$. The gap energies (between $-30$ and $+30$ meV) are integrated for clarity. The calculated scattering vectors from the interacting $k \cdot p$ band structure and the measured CDW wavevector are indicated with white dashed lines and the white shaded region respectively. STS parameters: $I_\text{set} = 1$ nA, $V_\text{mod} = 2.3$ mV, $f_\text{mod} = 736$ Hz, \textbf{a,b.} $V_\text{set} = 100$ mV, \textbf{e.} $300$ mV.}
  \label{fig4}
\end{figure*}

\paragraph*{}
\autoref{fig2}g and h summarize our FT-STS observations in the insulating interior, and near the topological edge of WTe$_2$, respectively. These figures are repeated in \autoref{fig3}b and \autoref{fig4}e, respectively, where they are discussed in greater detail.\autoref{fig2}g shows the FT-STS spectrum along $\Gamma$--X in the insulating interior which reveals the expected dispersing QPI features of the electron and hole pockets, separated by the bandgap. The electron pocket shows minima at $q \approx \pm 2 \Lambda$, reflecting inter-pocket scattering at approximately twice the wavevector $q = 2k$. Intra-pocket scattering of the electron pockets is also seen at low $q$. The single valence band at $\Gamma$ shows only intra-pocket scattering at $q = 2k$ as expected. These are confirmed by the overlaid calculated scattering vectors from the interacting $k \cdot p$ band structure (white dashed lines, see Experimental). An FT-STS spectrum acquired along the line in~\autoref{fig2}d which includes the topological edge state (\autoref{fig2}h) displays similar dispersing QPI features. Interestingly, additional modulations of the LDOS are observed within the gap at wavevector $q_\mathrm{CDW} = 0.29 \pm 0.01[2\pi/a]$, which are non-dispersive. The existence of LDOS modulations within the gap is inconsistent with QPI as an origin. We instead interpret these LDOS modulations as the signature of the CDW order parameter of the exciton condensate, imprinted on the topological edge state, and we explore these features in more detail.

\paragraph*{}
First, we use QPI features in FT-STS to explore the nature of the gap-collapse transition upon doping, and confirm the origin of the bandgap in the insulating state due to strong Coulomb interactions renormalising the band dispersion.~\autoref{fig3} shows FT-STS spectra in the insulating interior over different conditions of charge transfer doping. The insulating state (at low $V_G = 10$ V, \autoref{fig3}b, identical to \autoref{fig2}g) shows the expected band dispersion in the bulk, with the valence and conduction bands well separated by an insulating gap of approximately 60 meV. As previously shown~\cite{Bent2024_WTe2_EI_Gating} significant doping with holes or electrons induces a quantum phase transition, indicated by the collapse of the insulating gap. Here, we provide additional clear evidence for this collapse from QPI features in our gated FT-STS experiments. In \textit{p}-doped WTe$_2$ at $V_G = -40$ V (\autoref{fig3}a) the band edges appear to overlap at the charge neutrality point (CNP). A small Coulomb gap (CG) remains centred at the Fermi energy consistent with early reports of as-grown semimetallic WTe$_2$~\cite{Jia2017_WTe2_QPI_STM, Song2018_WTe2_QPI_STM}. The CNP and CG are reflected in the accompanying integrated LDOS spectrum. A similar effect is observed for significant electron doping ($V_G = +70$ V, \autoref{fig3}c), inducing a \textit{n}-doped semimetallic band structure. QPI confirms that the \textit{n}- and \textit{p}-doped states are both semimetallic, \textit{via} the clear observation of the coexistence of electron and hole pockets near the Fermi energy in \autoref{fig3}a,c. This is reflected in the integrated LDOS, where the excitonic gap collapses to the small CG and the CNP aligns with the meeting of the bands. The calculated scattering vectors from the interacting $k \cdot p$ band structure align very well with the band positions. Non-dispersing LDOS modulation at $q_\mathrm{CDW}$ were not observed in these spectra, acquired in the insulating 2D bulk. 

\paragraph*{}
\autoref{fig4} examines the FT-STS of WTe$_2$ near the topological edge of the sample in more detail.~\autoref{fig4}a,b shows the real-space (\autoref{fig4}a) and Fourier-space (\autoref{fig4}b) variation of LDOS at various energies within the bulk gap (between $-30$ and $+30$ meV), taken along the line shown in \autoref{fig2}d, and acquired at a perpendicular magnetic field of strength 3 T. The resolution of the CDW Fourier peaks is primarily limited to the decay length of the edge state into the bulk. In addition to the exponential decay of the edge LDOS into the interior expected for the topological edge mode, we observe LDOS oscillations which appear to be modulations of the topological edge-derived LDOS (their roughly constant amplitude on the vertical logarithmic scale indicates the oscillation amplitude is proportional to the exponentially decaying LDOS of the topological edge).~\autoref{fig4}b shows, remarkably, that the modulations of the LDOS are non-dispersive, with a mean frequency of $q_\mathrm{CDW} = 0.29 \pm 0.01[2\pi/a]$.~\autoref{fig4}c,d feature complementary theoretical calculations of the edge LDOS (at $E = 3.5$ mV) and corresponding Fourier transform respectively for the three excitonic phases (CDW/SDW, $\mathcal{T}$-CDW, and spin spiral). The edge LDOS shows good qualitative agreement with the CDW/SDW phase, and the measured wavevector $q_\mathrm{CDW}$ is consistent with the expected SDW/CDW order parameter of $2q_c \sim 0.31[2\pi/a]$~\cite{Kwan_SSvsSDW_2020,Crommie2017_WTe2_ARPES,Cucchi_WTe2_ARPES_2019}. These features cannot be a result of QPI or Friedel oscillations as both these phenomena rely on the scattering of charges across constant energy contours in the band structure~\cite{hawkes_QPI_definition_2019,hoffman_imaging_2002}. Particularly, the latter is dispersive~\cite{crommie_Friedel_1993,kanisawa_Friedel_2001} and only observed for systems with metallic Fermi surfaces~\cite{friedel_Original_1952} which fundamentally conflicts with the insulating state observed in the interior.

\paragraph*{}
\autoref{fig4}e (identical to \autoref{fig2}h) shows the FT-STS spectrum for the same line shown in \autoref{fig2}d, over a wider range of energies outside the gap. Similar to \autoref{fig3}b, the electron and hole pocket intra-pocket scattering vectors are well resolved, however the electron pocket inter-pocket scattering is less visible in this measurement. These features align well with those calculated from the interacting $k \cdot p$ band structure  (see supplementary information for details). Additional non-dispersive features at $q \simeq \pm 0.29 [2\pi/a]$ are observed inside the integrated in-gap spectrum, matching those seen in \autoref{fig4}b. Whether the non-dispersive feature persists at positive energies is difficult to determine due to the strong inter-pocket scattering features at $q \approx 2\Lambda$ associated with the electron pockets.

\section{Discussion}

\paragraph*{}
The measured CDW order parameter $2q_c = 0.29 \pm 0.01[2\pi/a]$ aligns well with experimentally derived values of $\Lambda$ from ARPES measurements \textendash~namely $2\Lambda \approx 0.3$~\cite{Crommie2017_WTe2_ARPES}, and $0.33[2\pi/a]$~\cite{Cucchi_WTe2_ARPES_2019}. The overlaid calculated scattering vectors from the interacting $k \cdot p$ band structure appears to fit the dispersing QPI features of the electron and hole pockets well in the FT-STS spectra.         


\paragraph*{}
Naively, the CDW order parameter expected in the general case using a two-band model~\cite{Jerome1967_Excitonic_Insulator} is simply $q_c = \Lambda$, equal to the separation between the electron and hole pockets in momentum space. The excitonic ground state $\mathcal{T}$-CDW with order parameter of $q_c = \Lambda$ is predicted for WTe$_2$ when time-reversal symmetry preservation is enforced~\cite{Papaj_CDW_QPI_2024}, but it is found to have slightly higher energy than its symmetry breaking counterparts (SDW/CDW and spin spiral states)~\cite{Kwan_SSvsSDW_2020}. Calculations of the $\mathcal{T}$-CDW state~\cite{Papaj_CDW_QPI_2024} predict modulations of the edge LDOS with period $1/\Lambda$, producing non-dispersive features at $q = \pm \Lambda$ inside the bandgap. We detect no strong features at $q = \pm \Lambda$ in both the edge and interior regions, and the LDOS modulations visible in real-space exhibit a period clearly smaller than $1/\Lambda$.

\paragraph*{}
The topological properties of WTe$_2$ are expected to be sensitive to time-reversal breaking phenomena, potentially opening a gap in the edge state~\cite{Wang_breakdown_2023}. While the SDW/CDW and spin spiral phases locally break time-reversal symmetry, integrating over the period of the SDW and spiral ($1/\Lambda \sim$ 2 nm) results in no net magnetic moment~\cite{Papaj_CDW_QPI_2024}. We do not observe a clear signature of a gap at the edge as reflected in the intensity of the edge state LDOS modulations (\autoref{fig4}a), which are primarily independent of energy. Even in the presence of a $\pm 3$ T out-of-plane magnetic  field (see Figure S1), persistence of enhanced edge LDOS within the bandgap energies is observed, which is consistent with transport measurements~\cite{Fei_WTe2_Transport_2017} in which edge conduction persists. The culmination of this evidence strongly indicates that the topological edge state persists in the presence of an excitonic condensate in the interior, which coincides with theoretical predictions~\cite{Papaj_CDW_QPI_2024,Wang_breakdown_2023} for the SDW/CDW state. Interaction of the SDW/CDW and spin spiral state orders with the edge state is expected to open a small gap which is spatially periodic at $\pm \Lambda/2$~\cite{Wang_breakdown_2023}. We do not observe any suppression of LDOS due to formation of periodic gaps. This is possibly influenced by a combination of limited energy resolution and local disorder in the samples.

\paragraph*{}
The phase diagram of the two time-reversal symmetry breaking excitonic phases~\cite{Kwan_SSvsSDW_2020, parra-martinez_excitonic_phases_2025} predicts that the SDW/CDW phase occurs at a much smaller magnitude and range of Coulomb interaction strengths compared to the spin spiral phase. Our work utilises the conducting substrates graphene and graphite, expected to provide strong screening and reduced Coulomb interaction. The experiments in \autoref{fig4} use a graphite substrate; at low frequency and long wavelength graphite's screening is metallic, but at energies of 0.5--1.0 eV interband screening in graphite produces dielectric behaviour with $\varepsilon$ in the ranges of 14--40~\cite{philipp_Graphite_infrared_1977}, which is in reasonable agreement with the range of substrate dielectric constants (17--39) expected to produce the CDW/SDW regime in~\cite{Kwan_SSvsSDW_2020,parra-martinez_excitonic_phases_2025}. Transport experiments~\cite{Jia2021_WTe2_EI_Transport} on WTe$_2$ encapsulated by hexagonal boron nitride (h-BN) have been assumed to be in the spin spiral phase due to the low dielectric constant of h-BN ($\varepsilon_\text{h-BN} =$ 3.5~\cite{laturia_dielectric_hBN_2018}). This raises the possibility that transport experiments may not study the same phase as our (and other) spectroscopic measurements. To our knowledge, all spectroscopic (ARPES, STS) studies of monolayer WTe$_2$ have been performed on conducting substrates, with the exception of \citeauthor{Jia2021_WTe2_EI_Transport} who performed STS and planar tunnelling measurements on WTe$_2$ encapsulated with h-BN\@. Interestingly \citeauthor{Jia2021_WTe2_EI_Transport} may have observed a substantially larger bandgap, up to 109 meV, in STS of h-BN encapsulated WTe$_2$, potentially due to a different phase of the exciton condensate.

\paragraph*{}
The question remains as to why the CDW order parameter is not observed uniformly everywhere in the WTe$_2$ interior\@. At positive energies, the CDW wavevector is expected to overlap with the electron pocket inter-pocket scattering QPI feature located at $q \approx 2\Lambda$, making unambiguous observation of the CDW difficult at positive bias. At negative energies, a QPI feature due to interpocket scattering at $q \approx 2\Lambda$ is not expected nor observed, however, a non-disperive feature due to the CDW is absent also. The simplest explanation is that the effect of CDW on the bulk LDOS is simply too weak to be observed. The conducting topological edge state provides LDOS around the Fermi energy which can respond to the CDW potential providing a visible signature~\cite{Papaj_CDW_QPI_2024}, in contrast to the insulating bulk. We note that a non-dispersive feature at $q = \pm 1/4[2\pi/a]$ which overlaps with the hole pocket at negative energies is sometimes observed in the unfiltered data (see Figure S2), possibly indicating that this is an edge effect. Alternatively, the double derivative of these spectra (Figure S3) seem to indicate that this feature is dispersive and could provide evidence of band folding.


\paragraph{}
In summary, we have observed the symmetry-breaking order parameter of the exciton condensate ground state in monolayer WTe$_2$. The wavevector of the order parameter ($q = 0.29 \pm 0.01[2\pi/a]$) and the absence of gap opening at the edge (absence of strong time-reversal symmetry breaking) indicate that the order parameter is the predicted SDW/CDW order occurring at relatively weak interaction strength~\cite{Kwan_SSvsSDW_2020, parra-martinez_excitonic_phases_2025}, consistent with strong screening provided by the metallic (HOPG and graphene) substrates. We find that the visualisation of the order parameter in tunnelling experiments is intimately tied to the topological edge LDOS, in line with recent theoretical predictions~\cite{Papaj_CDW_QPI_2024}. This could provide a reason as to why it has not been captured until now, especially for studies utilising exfoliated monolayer WTe$_2$ samples.

\section{Experimental}

\paragraph*{}
WTe$_2$ sub-monolayer crystals were synthesised \textit{via} molecular beam epitaxy (MBE). Two different sample architectures were employed: exfoliated (monolayer) graphene on 300 nm SiO$_2$ over \textit{p}-type silicon, and HOPG, each grown in their respective system.

\paragraph*{MBE on Graphene over SiO$_2$/Si:}
Samples were prepared in an Omicron Lab10 ultra-high vacuum (UHV) MBE chamber~\cite{Jia_WTe2_STM_Luttinger2022} (base pressure $< 1 \times 10^{-10}$ mbar). Fabrication of the graphene/SiO$_2$/Si devices was completed in UHV and inert Ar atmosphere to facilitate clean and uncontaminated surfaces (see Ref.~\citenum{Bent2024_WTe2_EI_Gating} for details). MBE growth of WTe$_2$ crystals was carried out on graphene/SiO$_2$/Si substrates held at 160 $^\circ$C using co-deposition of W (99.998\%) and Te (99.999\%) with a flux ratio of 1:280 for 1 hour to achieve $\sim$ 40--50\% monolayer coverage.

\paragraph*{MBE on HOPG:}
HOPG bulk crystals (SPI supplies \textendash~grade A) were cleaved under nitrogen gas flow before being transferred to UHV (base pressure $< 2 \times 10^{-10}$ torr). The graphite substrates were degassed at 150 $^\circ$C. While maintaining this substrate temperature, MBE growth of WTe$_2$ was carried out \textit{via} e-beam evaporation (Focus GmbH) of W (99.998\%, flux $=$ 10--11 nA) and effusion (Kentax GmbH, dual cell) of Te (99.999\%, $T = 340$ $^\circ$C) for 30 minutes to achieve $\sim$ 90\% monolayer coverage.

\paragraph*{STM/STS:}
Samples grown on graphene over SiO$_2$/Si were measured in a Unisoku mK-USH1600 low-temperature STM (base temperature $\sim$ 30 mK, junction temperature $\sim$ 150 mK, base pressure $< 1 \times 10^{-10}$ mbar), while those grown on HOPG were measured in a SPECS Joule-Thomson low-temperature STM (base temperature $\sim$ 1.1 K, base pressure $< 1 \times 10^{-10}$ mbar). Both STM systems utilised chemically etched W or mechanically cut Pt/Ir tips, which were calibrated using the Au(111) Shockley surface state before spectroscopic measurements. Standard lock-in techniques were utilised for spectroscopic measurements. Modulation amplitudes and frequencies can be found in accompanying figure descriptions.

\paragraph*{Fourier transform STS data processing:}
Fourier transform STS spectra were obtained by taking the Fourier transform of LDOS spectra acquired along the $x$(short)-axis of WTe$_2$. The LDOS spectra were first treated with a Hann window function to reduce edge effects, which retains enough spectral weight around the topological edge to resolve the CDW order parameter. When a Fourier transform is applied, this window acts visually similar to taking the derivative in both Fourier space axes (double derivative), a common technique employed in ARPES data analysis to better visualise dispersive bands~\cite{Li_QPI_Hann_2020,harris_HannWindow_1978}, while also removing spurious edge effects. True numerical derivatives in the energy axis have the effect of diminishing non-dispersive features and so were avoided in this study. Figure S2 compares untreated and double derivative and Hann windowed FT-STS spectra. No symmetrisation of the resultant Fourier transforms was performed. The intensity of the integrated gap LDOS for FT-STS spectra of the edge regions is artificially enhanced to be visualised on the same colour scale. Figure S4 shows the same data without the integrated gap LDOS, and are visualised on the same colour scale using log-scale intensity. The value of the CDW order parameter was calculated by performing Gaussian fits to the constant energy cuts in~\autoref{fig4}b, where we quote the mean of the fits to one standard deviation. All analyses were performed in Python using NumPy~\cite{harris_nparray_2020}, SciPy~\cite{virtanen_scipy_2020} and Matplotlib~\cite{hunter_matplotlib_2007}.

\paragraph*{Interacting $\boldsymbol{k \cdot p}$ band structure:}
The electronic structure of WTe$_2$ is described with a $k \cdot p$ model~\cite{Kwan_SSvsSDW_2020,Bent2024_WTe2_EI_Gating,Jia2021_WTe2_EI_Transport}. Near
the $\Gamma$ point, the Hamiltonian of a four-band model reads:

\begin{multline}
    \hat{h}(k) = \varepsilon_{+}(k) + \left[\varepsilon_{-}(k)+\delta\right] \tau^z \\
    + v_x k_x \tau_x s_y + v_y k_y \tau_y s_0
\end{multline}
where $\tau^\mu$ and $s^\mu$ are Pauli matrices, representing orbital and spin
degree of freedom. $\tau^z = \pm 1$ refers to \textit{d} and \textit{p} orbitals, respectively. The parameters used can be found in Ref~\citenum{Kwan_SSvsSDW_2020}. With this set of parameters, the conduction band minimum and the valence band maximum are connected by a wave vector of length $q_c = \frac{1}{6}[2\pi/a]$. The Coulomb interactions in the charge density wave phase are considered by the interacting Hamiltonian:

\begin{equation}
    H_{\mathrm{int}}=\frac{1}{2 N_k} \sum_{\boldsymbol{k}, \boldsymbol{p}, \boldsymbol{q}, \alpha, \beta} U(\boldsymbol{q}) c_{\boldsymbol{k}+\boldsymbol{q}, \alpha}^{+} c_{\boldsymbol{p}-\boldsymbol{q}, \beta}^{+} c_{\boldsymbol{p}, \beta} c_{\boldsymbol{q}, \alpha}
\end{equation}
where $c_{\boldsymbol{k}, \alpha}^{+}$ ($c_{\boldsymbol{k}, \alpha}$) are the creation (annhiliation) operators for an electron with momentum $\boldsymbol{k}$ and orbit index $\alpha$, and $U(\boldsymbol{q}) = \frac{2U_0}{q \xi} \tanh\frac{q \xi}{2}$ is a model screened interaction with a screening length $\xi =$ 25 nm. The value of $U_0$ is chosen to match the gap from the self-consistent calculations
with the gap observed in the experiments. To simulate the charge doping effect,
we introduce a chemical potential term $\mu$ in the non-interacting part of the
Hamiltonian.

\paragraph*{Real space calculations of excitonic insulator edge LDOS:}
The modelling of the impact of the excitonic condensate on the topological edge state was performed using a lattice-discretized version of the Bernevig-Hughes-Zhang model~\cite{bernevig_BHZModel_2006}, with the direction of the spin-orbit coupling term taken as in the WTe$_2$ model above. This choice of model is motivated by the fact that the $k\cdot p$ model which reproduces the low-energy band dispersion of WTe$_2$ tends to greatly underestimate the penetration depth of the topological edge states ($l < 1$ nm) as compared to the experiments, where $l \sim 3.6$ nm. The momentum space form of this Hamiltonian is:
\begin{multline}
    H(\mathbf{k}) = A (k_x \tau_x s_y + k_y \tau_y s_0)  \\ 
    +  (M + B k^2) \tau_z s_0 + D k^2 \tau_0 s_0
\end{multline}
where $k^2=k_x^2+k_y^2$ and $\tau_i$, $s_i$ are Pauli matrices representing orbital and spin degrees of freedom. The model is then discretized on a square lattice with a lattice constant $a=1\, \text{\AA}$ using the standard finite difference method. Calculations are performed for a ribbon of width $W=500\,\text{\AA}$, which is infinite along the $x$ direction (allowing the use of translational invariance), and the hard-wall boundary conditions are chosen in the $y$ direction.

Comparisons are made between 3 different order parameters, which correspond to the results obtained through the Hartree-Fock calculations in Refs.~\citenum{Kwan_SSvsSDW_2020,Wang_breakdown_2023}: time-reversal-preserving charge density wave (CDW), time-reversal-breaking spin density wave (SDW), and spin spiral. Each of these order parameters makes a different contribution to the real space Hamiltonian of the system:
\begin{equation}
    H_\text{CDW}(\mathbf{r}) = \Delta_\text{CDW} \cos\left(q_c y\right) \tau_0 s_0
\end{equation}

\begin{equation}
    H_\text{SDW}(\mathbf{r}) = \Delta_\text{SDW} \sin\left(q_c y\right) \tau_0 (s_x + s_z)
\end{equation}

\begin{multline}
    H_\text{spiral}(\mathbf{r}) = \Delta_\text{spiral} (\cos\left(q_c y\right) \tau_0 s_x \\ 
    + \sin\left(q_c y\right) \tau_0 s_z)
\end{multline}
where $q_c = 0.26\,\text{\AA}^{-1}$ is the momentum separation between electron and hole pockets as found from the experimental determination of the band dispersion. The modulation direction is perpendicular to the edge, so it does not affect translational invariance along the $x$ direction. Local density of states is then calculated at energy $E=3.5$ meV for each of the order parameters separately and plotted in the vicinity of the edge of the ribbon, resulting in data presented in~\autoref{fig3}. In performing the calculations, KWANT package was used~\cite{groth_kwant_2014}. The parameter values are: $A=0.25 \,\text{eV \AA}$, $B=-10 \,\text{eV \AA}^2$, $D=-8 \,\text{eV \AA}^2$, $M=-5$ meV, $\Delta_\text{CDW}=5$ meV, $\Delta_\text{SDW}=22$ meV, $\Delta_\text{spiral}=20$ meV. The model parameters are chosen such that the penetration depth of the topological edge state is about 36 \AA\, when no order parameter is present, which is the penetration depth seen in the experiment.

\begin{acknowledgement}

  I.D.B.~acknowledges support from MSCA Program (101063547-GAP-101063547). M.T.E.~acknowledges funding support from ARC Future Fellowship (FT2201000290). L.W., I.D.B., M.T.E. and M.S.F. acknowledge funding from the FLEET Centre of Excellence, ARC Grant No. CE170100039. IMDEA Nanociencia and IFIMAC acknowledge financial support from the Spanish Ministry of Science and Innovation `Severo Ochoa' (Grant CEX2020001039-S) and `Mar\'{i}a de Maeztu' (Grant CEX2018000805-M) Programme for Centers of Excellence in R\&D, respectively. B. W. acknowledges the support of the National Research Foundation (NRF) Singapore, under the Competitive Research Program ``Towards On-Chip Topological Quantum Devices'' (NRF-CRP21-2018-0001), with further support from the Singapore Ministry of Education (MOE) Academic Research Fund Tier 3 grant (MOE-MOET32023-0003) “Quantum Geometric Advantage”. M.G. has received financial support through the “Ram\'{o}n y Cajal”
  Fellowship program (RYC2020-029317-I), “Ayudas para Incentivar la
  Consolidación Investigadora” (CNS2022-135175), and the Spanish
  Ministerio de Ciencia e Innovación (Grant no. PID2021-123776NB-C21).
  Financial support through the (MAD2D-CM)-MRR MATERIALES
  AVANZADOS-IMDEA-NC and (MAD2D-CM)-MRR MATERIALES AVANZADOS-UAM is also
  acknowledged.

\end{acknowledgement}

\begin{suppinfo}

Additional FT-STS data processing method details, figures of alternate data processing, and comparison between data acquired with and without out-of-plane magnetic field ($B_\perp =$ 3 T). (PDF).
\begin{itemize}
  \item WTe2OrderParamSuppInfo.pdf
\end{itemize}

\end{suppinfo}

\bibliography{bibliography}

\end{document}